\documentclass[11pt]{article}
\usepackage{moriond,epsfig}

\def\be{\begin{equation}}
\def\ee{\end{equation}}
\def\bea{\begin{eqnarray}}
\def\eea{\end{eqnarray}}

\begin{document}
\vspace*{1cm}
\title{Long range phenomena in heavy­ ion collisions observed by the PHOBOS experiment}

\author{ Krzysztof Wo\'{z}niakfor the PHOBOS Collaboration
\footnote{
%
%
B.Alver$^4$,
B.B.Back$^1$,
M.D.Baker$^2$,
M.Ballintijn$^4$,
D.S.Barton$^2$,   
R.R.Betts$^6$,
A.A.Bickley$^7$,
R.Bindel$^7$,
W.Busza$^4$,
R.Bindel$^7$,
W.Busza$^4$,
A.Carroll$^2$,
Z.Chai$^2$,
V.Chetluru$^6$,
M.P.Decowski$^4$,  
E.Garc\'{\i}a$^6$,
T.Gburek$^3$,
N.George$^2$,
K.Gulbrandsen$^4$,
C.Halliwell$^6$,
J.Hamblen$^8$,
I.Harnarine$^6$,  
M.Hauer$^2$,
C.Henderson$^4$,
D.J.Hofman$^6$,
R.S.Hollis$^6$,
R.Ho\l y\'{n}ski$^3$,
B.Holzman$^2$,
A.Iordanova$^6$,  
E.Johnson$^8$,
J.L.Kane$^4$,
N.Khan$^8$,
P.Kulinich$^4$,
C.M.Kuo$^5$,    
W.Li$^4$,
W.T.Lin$^5$,
C.Loizides$^4$,
S.Manly$^8$,
A.C.Mignerey$^7$,   
R.Nouicer$^2$,
A.Olszewski$^3$,
R.Pak$^2$,
A.Olszewski$^3$,
R.Pak$^2$,
C.Reed$^4$,
E.Richardson$^7$,   
C.Roland$^4$,
G.Roland$^4$,
J.Sagerer$^6$,
H.Seals$^2$,
I.Sedykh$^2$,   
C.E.Smith$^6$,
M.A.Stankiewicz$^2$,
P.Steinberg$^2$,
G.S.F.Stephans$^4$,
A.Sukhanov$^2$,
A.Szostak$^2$,  
M.B.Tonjes$^7$,
A.Trzupek$^3$,
C.Vale$^4$,
G.J.van~Nieuwenhuizen$^4$,
S.S.Vaurynovich$^4$,
R.Verdier$^4$,    
G.I.Veres$^4$,
P.Walters$^8$,
E.Wenger$^4$,
D.Willhelm$^7$,
F.L.H.Wolfs$^8$,  
B.Wosiek$^3$,
 K.Wo\'{z}niak$^3$,
S.Wyngaardt$^2$,
B.Wys\l ouch$^4$ \\
$^1$~Argonne National Laboratory, Argonne, IL 60439-4843, USA\\
$^2$~Brookhaven National Laboratory, Upton, NY 11973-5000, USA\\
$^3$~Institute of Nuclear Physics PAN, Krak\'{o}w, Poland\\
$^4$~Massachusetts Institute of Technology, Cambridge, MA 02139-4307, USA\\
$^5$~National Central University, Chung-Li, Taiwan\\
$^6$~University of Illinois at Chicago, Chicago, IL 60607-7059, USA\\
$^7$~University of Maryland, College Park, MD 20742, USA\\
$^8$~University of Rochester, Rochester, NY 14627, USA
 }
}

\address{Institute of Nuclear Physics PAN, 
Krak\'{o}w, Poland}

\maketitle\abstracts{
The PHOBOS experiment at RHIC has measured large samples of Au+Au and Cu+Cu
collisions using a detector with uniquely large angular acceptance.  
These data enable studies of particle production over a very wide pseudorapidity
interval which reveal unexpected features.
In the analysis of correlations with a high-$p_{_{T}}$ trigger particle ($p_{_{T}}>2.5$ GeV/c)
a ridge extending at least 4 units of pseudorapidity was found.
The results on forward-backward and two-particle correlations suggest
that particles are produced in very large clusters which are wider
in pseudorapidity than is expected for isotropic decays.
Explanation of these experimental results requires models in which
both short-range and long-range correlations are present. 
}

Since the year 2000 at the Relativistic Heavy Ion Collider (RHIC), collisions of heavy ions at the highest energies have been  
measured 
and analysed. Several phenomena found in these studies manifest 
creation of a new type of matter, strongly interacting
Quark-Gluon Plasma (sQGP). Most noticeable are the absorption of partons
observed as suppression of high-$p_{_{T}}$ particles or jets and 
the collective effects visible as a strong elliptic flow
\cite{whitepaper}. 
Better understanding of properties of the matter created 
in heavy ion collisions and a search for signs of  potential 
phase transition require detailed analysis of many observables 
and studies of correlations between them.

The PHOBOS experiment measured all types of collisions available 
at RHIC using a detector optimized for registering charged particles2
in almost full solid angle - with the multiplicity detector
covering uniquely wide range $|\eta|<5.4$. 
In the spectrometer, momenta of about 1\% of charged particles were 
determined. Using this detector it was possible to measure 
the yields of particles with extremally small transverse momenta 
(starting from 30~MeV/c for pions). 
The comparison of yields of charged particles 
at high-$p_{_{T}}$ in  
$d+Au$ and $Au+Au$ collisions clearly shows that partons created 
in hard scattering of quarks or gluons interact in the dense
matter created in the central $Au+Au$ collisions. 
Usually at least one of back-to-back emitted partons is stopped 
and only a jet or a high-$p_{_{T}}$ particle originating from
the second parton is registered. 

The analysis \cite{pho_prl104_correl_trig} of the correlations
between a trigger particle with $p_{_{T}}>2.5$ GeV/c and 
other charged particles as 
a function of  the difference
of pseudorapidity and azimuthal angle, $\Delta\eta$ and $\Delta\phi$, allows to study the interaction 
of the stopped parton with the medium.
In the central $Au+Au$ collisions the yield of particles
correlated with the high-$p_{_{T}}$ trigger particle is larger
than in the $p+p$ interactions. In the {\it near side}, 
$|\Delta\phi| \approx 0 \pm 1$, a ridge extending up to the end 
of acceptance range ($-4<\Delta\eta<2$) is present. It can be
described as an additional yield which adds uniformly in $\eta$
to the yield observed in $p+p$ interactions. 
In the {\it away side}, $\Delta\phi \approx \pi \pm 2$,
such additional yield is even larger. 
Analysis of these yields as a function of centrality
shows that they are the largest in central 
collisions and decrease for peripheral collisions.
At $N_{part} \approx 80$ the difference between $Au+Au$ and $p+p$
for the {\it near side} drops to zero.

\begin{figure}
\begin{minipage}{6cm}
\psfig{figure=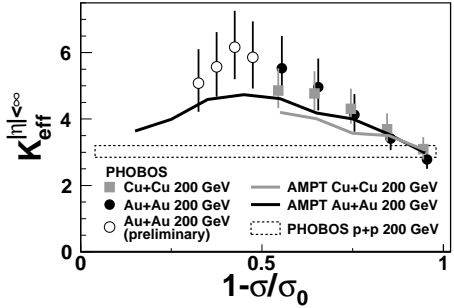,width=6cm}
\end{minipage}
\hspace{0.6cm}
\begin{minipage}{9cm}
\caption{The effective size of the clusters corrected for the acceptance effects,
as a function of the fractional cross section, i.e. obtained for similar
geometry of $Au+Au$ and $Cu+Cu$ collisions. The cluster parameters
obtained for elementary $p+p$ collisions and the results of the same 
reconstruction procedure performed on the events from AMPT generator 
are shown for comparison.
\label{fig:clusterparameters_fract_keff}}
\end{minipage}
\end{figure}

\begin{figure}[htb]
\begin{center}
\psfig{figure=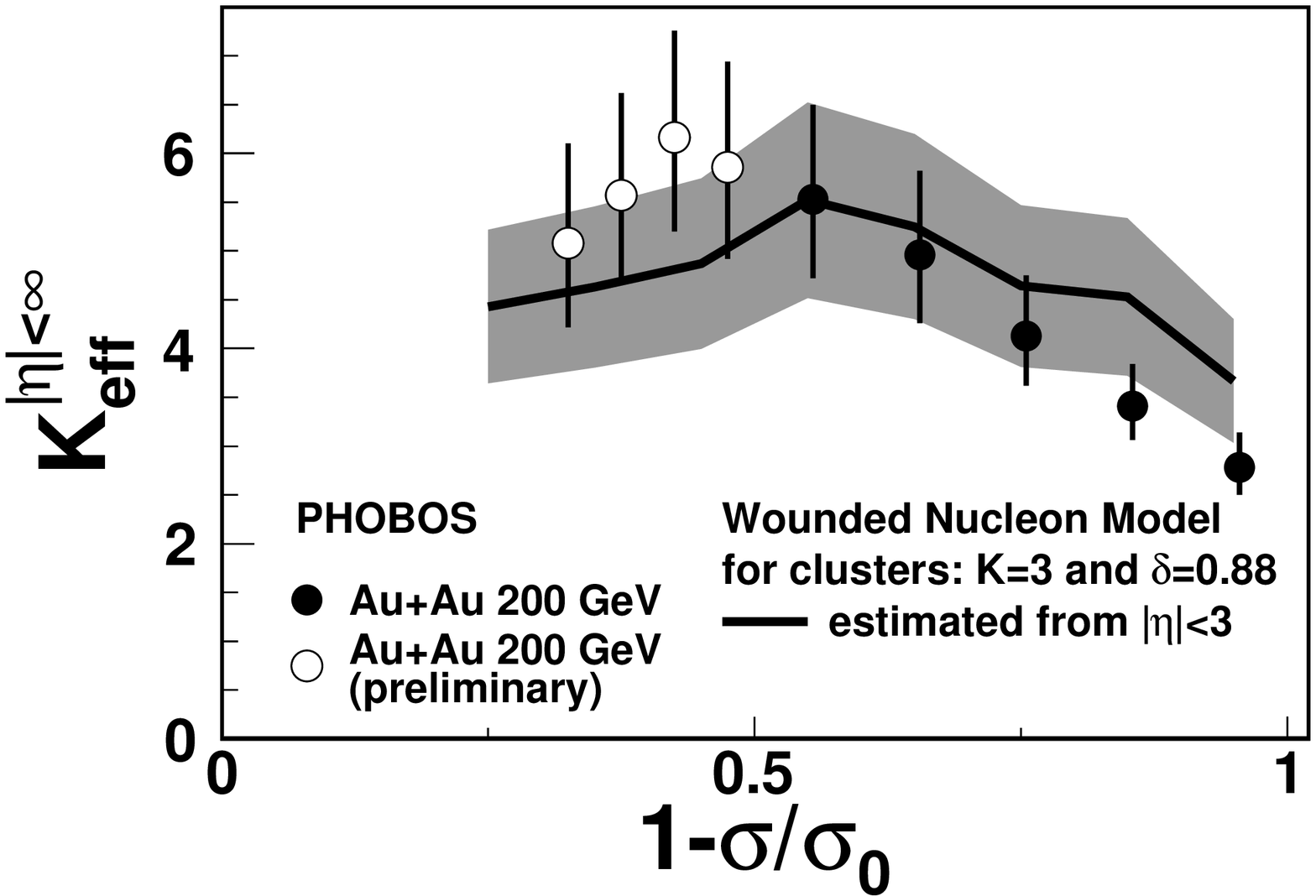,width=6cm}
 \hspace{0.6cm}
\psfig{figure=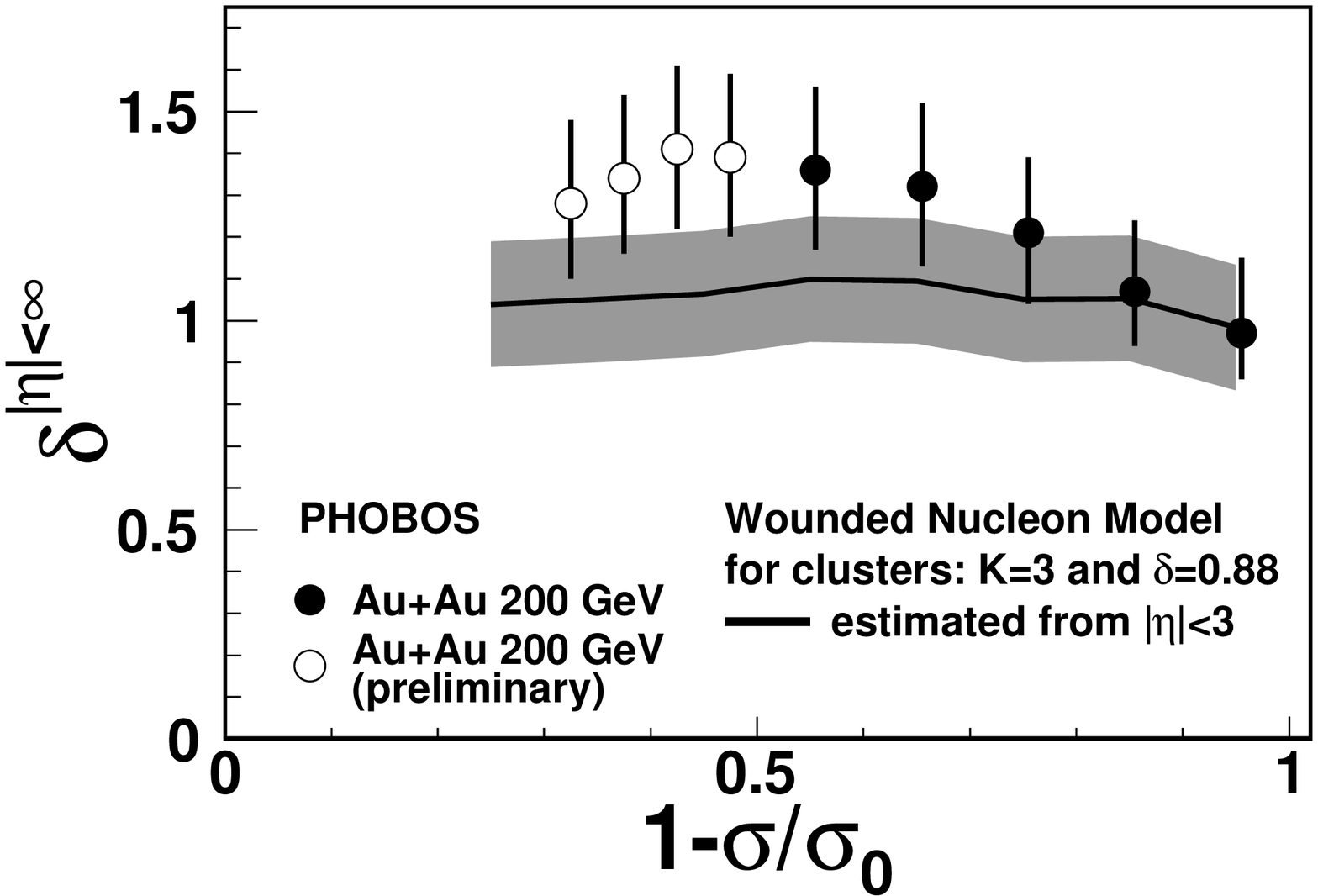,width=6cm}
\end{center} 
\caption{The effective cluster size (left) and the width parameter $\delta$ (right) for $Au+Au$ collisions
compared with the predictions from the Wounded Nucleon Model
in which we assume fragmentation of the wounded nucleons 
into clusters similar 
to those observed in $p+p$ interactions.
\label{fig:clusterparameters_fract_wounded}}
\end{figure}

The correlations between all charged particles registered in the 
PHOBOS multiplicity detector
were studied in a very wide range, $|\eta|<3$, for
elementary $p+p$ interactions (at 200 GeV and 410 GeV
\cite{pho_prc75_correl_pp_cluster})
and nuclei collisions, $Cu+Cu$ and $Au+Au$ at 200 GeV
\cite{pho_prc81_correl_2part_CuCu_AuAu,pho_qm2009_correl_2part_CuCu_AuAu}.
They are represented by a correlation function: \\
\hspace*{2cm} $R(\Delta \eta,\Delta \phi)=\left<(n-1)\left(\frac{\rho_{n}^{\rm II}
(\Delta \eta,\Delta \phi)}{\rho^{\rm mixed}(\Delta \eta,\Delta \phi)}-1\right)\right> $ \\
where $\rho(\eta_{1},\eta_{2},\phi_{1},\phi_{2})$ is 
the charged pair density distribution for measured 
events (in the numerator)
or for uncorrelated pairs taken from different events 
(in the denominator). In the further analysis the function
integrated over one of the variables, $R(\Delta\eta)$ or 
$R(\Delta\phi)$, is used. The first of them has a maximum
at $\eta \approx 0$ which is expected for short range correlations.
It is thus natural to describe particle production
as a two-step process: production of some intermediate 
objects, clusters, which then decay into finally observed
particles \cite{cluster_model}.
Using the correlation function  $R(\Delta\eta)$ it is possible
to extract the parameters of the clusters: $K_{eff}$, the effective
cluster size, and $\delta$, width of the two-particle correlation.
It is worth to note that even for very large acceptance, 
six pseudorapidity units, available in PHOBOS,
acceptance corrections are large.  
Already in the elementary interactions\cite{pho_prc75_correl_pp_cluster} 
the cluster size is large, $K_{eff} \approx 3$. Even larger
values, up to $K_{eff} \approx 6$, are found in nuclei collisions \cite{pho_prc81_correl_2part_CuCu_AuAu,pho_qm2009_correl_2part_CuCu_AuAu},
as shown in Fig.\ref{fig:clusterparameters_fract_keff}.
The width of the clusters (shown later 
in  Fig.\ref{fig:clusterparameters_fract_wounded}) is also large, 
exceeding that expected for isotropic decay at rest.
We observe the same cluster parameters for 
$Cu+Cu$ and $Au+Au$ collisions with 
similar geometries even if the number of nucleons participating in the collisions and total multiplicities are much different
\cite{pho_prc81_correl_2part_CuCu_AuAu,pho_qm2009_correl_2part_CuCu_AuAu}.

\begin{figure}[tb]
\begin{center}
\psfig{figure=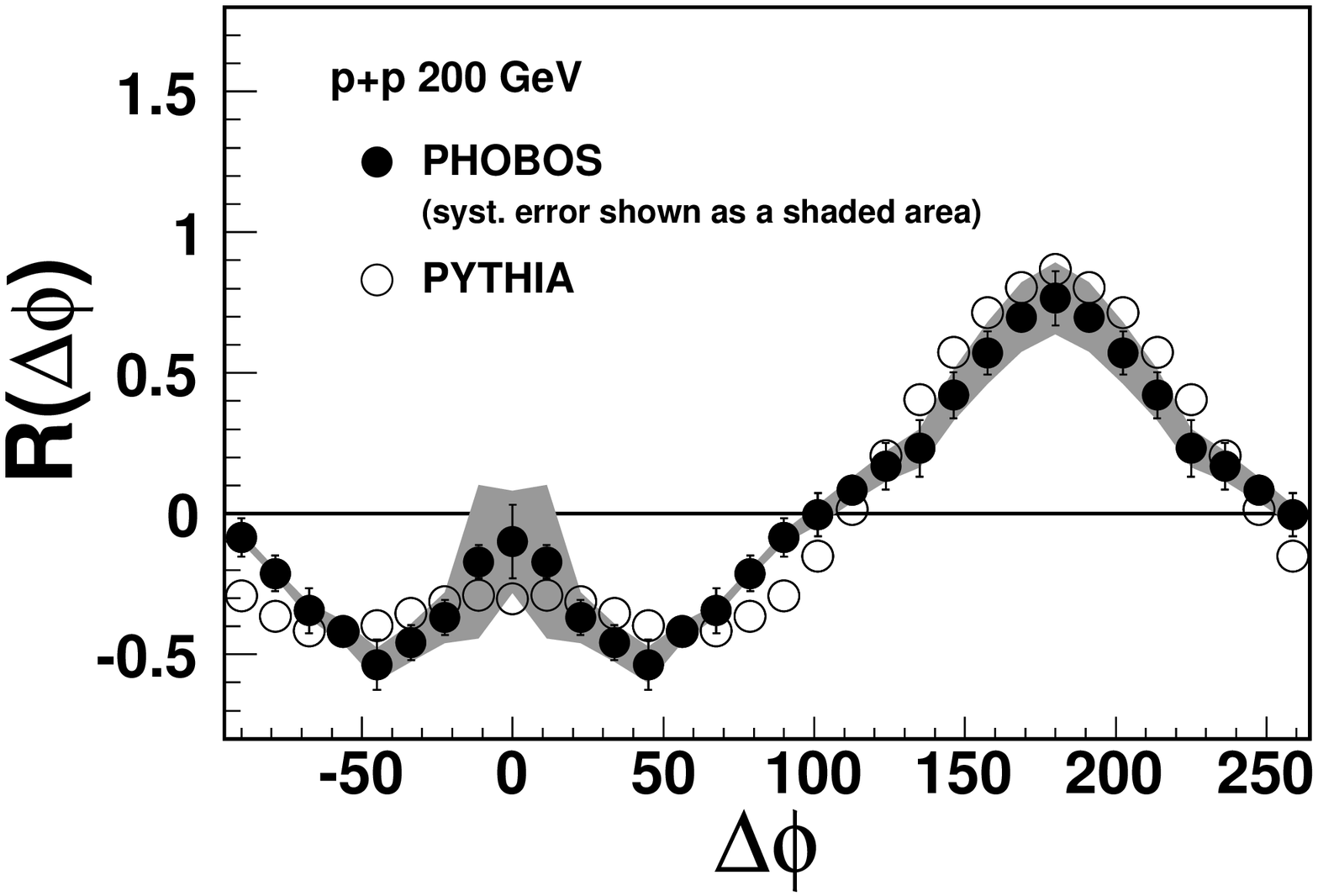,width=6cm}\hspace{0.6cm}
\psfig{figure=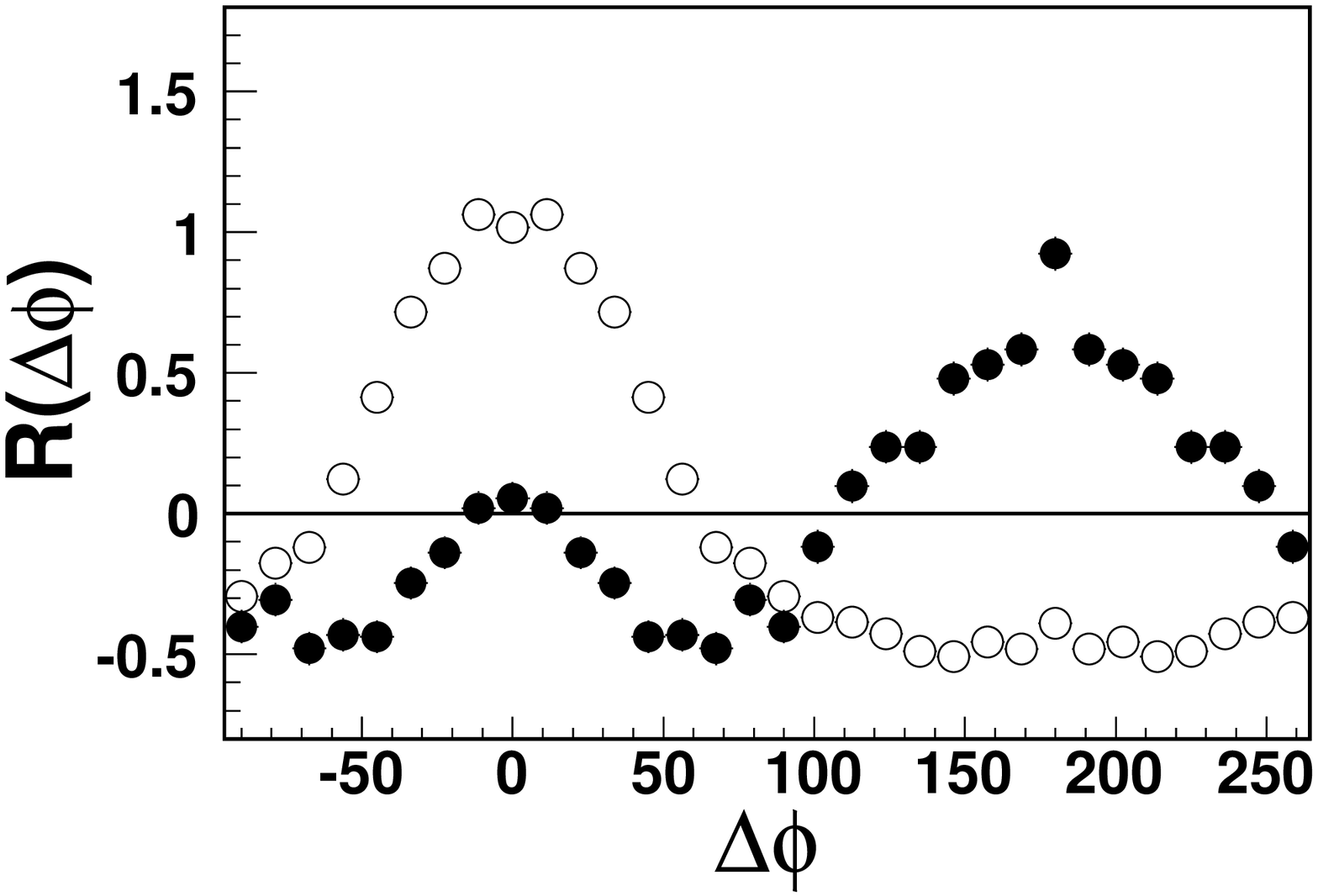,width=6cm}
\end{center}
\caption{Two-particle correlation functions $R(\Delta\phi)$ 
obtained for the  experimental data (left) and for generated events
with identical clusters (decays: $\omega(782) \rightarrow \pi^{+}\pi^{-}\pi^{0}$) (right).
In the right plot the correlation for events with randomly 
generated clusters (open circles) 
and modified so that sum of all transverse momenta 
is zero (full circles) are compared.}
\label{fig:pp_2part_dphi}
\end{figure}

Also, the correlation function $R(\Delta\phi)$ 
can be explained
by the cluster model. However, a simple assumption 
that clusters' momenta can be 
randomly generated from a universal function reproducing
only global $dN/d\eta$ and $dN/dp_{_{T}}$ distributions leads
to a shape totally different from that measured experimentally.
An agreement can be achieved only after 
enforcing transverse momentum conservation 
(by slightly modifying momenta to ensure $\Sigma p_{_{T}} = 0$)
as can be seen in Fig.\ref{fig:pp_2part_dphi}.

\begin{figure}[b]
\begin{minipage}{11cm}

\psfig{figure=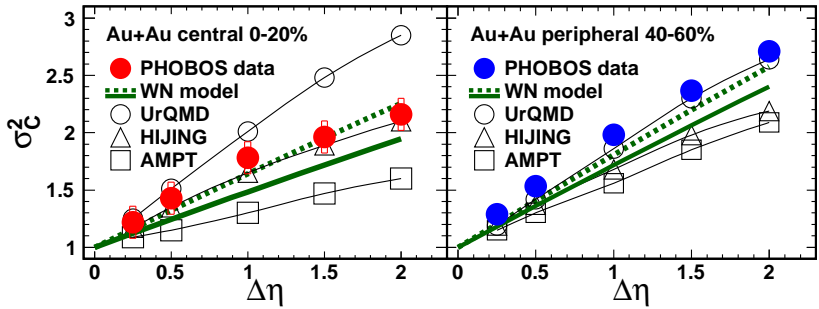,width=11cm}
\end{minipage}
\hspace{0.2cm}
\begin{minipage}{4.4cm}
\caption{The values of $\sigma_{C}^{2}$ as a function
of pseudorapidity bin width $\Delta\eta$ for central and peripheral
$Au+Au$  collisions at $\sqrt{s_{_{NN}}}=200$ GeV.
Experimental results
are compared with predictions from several 
models.}
\label{fig:sigc_wounded}
\end{minipage}
\end{figure}

The large acceptance of the PHOBOS detector allows to measure
forward-backward correlations at large distances. In this study, 
an asymmetry variable $C = (N_{B}-N_{F})\sqrt{N_{F}+N_{B}}$
is used, where
$N_{B}$ and $N_{F}$ denote the number of charged particles measured
in two pseudorapidity bins, symmetric with respect to $\eta=0$,
at negative and positive $\eta$, respectively \cite{pho_prc74_fb_correl}. 
This variable is insensitive to the dependence of total
multiplicity on centrality of the collision (that is enforcing
$\langle N_{B}\rangle = \langle N_{F}\rangle$), but the variance $\sigma_{C}^{2}$
measures the strength of multiplicity fluctuations.
For purely statistical fluctuations we obtain $\sigma_{C}^{2}=1$.
The PHOBOS Collaboration has measured $\sigma_{C}^{2}$ 
as a function of the width of pseudorapidity bin, $\Delta\eta$, 
and the position of bin center, $\eta$.  
$\sigma_{C}^{2}$~is~larger 
than~1 and increases when $\eta$ and especially $\Delta\eta$
increase \cite{pho_prc74_fb_correl}. 
This agrees qualitatively with the expectation from
the cluster model, in which the observed dependencies
are explained by acceptance effects. However, in this case it
is not possible to extract both parameters of the clusters at 
the same time, as for large and wide clusters these 
dependencies may look similar as for small but narrow clusters
\cite{pho_ijmpe16_rare_fb_qm2006}.
Unexpectedly, but consistently with the trends observed 
for 2-particle correlations, the values of $\sigma_{C}^{2}$ 
found for central $Au+Au$  collisions
are smaller than for peripheral collisions 
(see Fig.\ref{fig:sigc_wounded}), indicating 
different effective cluster sizes (and possibly 
also $\delta$ width).

The centrality dependence of $\sigma_{C}^{2}$ values, 
shown in Fig.\ref{fig:sigc_wounded}, is
not well reproduced by the models of particle production: 
UrQMD \cite{urqmd} has wrong centrality 
dependence, HIJING\cite{hijing} gives in both cases the same values 
and AMPT\cite{ampt}
predicts correct trend, but underestimates $\sigma_{C}^{2}$.
The best agreement\cite{bzdak_wozniak_wounded_sigmac} is observed for 
the Wounded Nucleon Model\cite{WNM_def}
which assumes that the nucleons taking part in the 
collision are the source of the particles which are produced 
according to a universal fragmentation function,
asymmetric in $\eta$ and about 10 pseudorapidity
units wide, extracted 
from the data on $d+Au$ collisions. The fluctuations
observed as large values of $\sigma_{C}^{2}$ are a sum 
of these present already in $p+p$ interactions (possibly from
production of clusters) and those generated by 
the fluctuations of the number of wounded nucleons,
in forward and backward moving nuclei\cite{bzdak_wozniak_wounded_sigmac}. 

The Wounded Nucleon Model may be used to describe not only 
forward-backward fluctuations, but also 2-particle correlations.
In this case, predictions are less precise, as it is more difficult
to include short range correlations (from $p+p$). Obviously,2 
wounded nucleons should fragment into clusters
which then decay into final particles. As an approximation
identical clusters with  $K_{eff}=3$ and $\delta=0.88$,
effective parameters found in $p+p$ interactions
\cite{pho_prc75_correl_pp_cluster}, can be used.
The fragmentation 
function into clusters should have similar shape
(but $K_{eff}$ times smaller integral) 
as for fragmentation into particles.
These assumptions allow to obtain  Wounded Nucleon Model predictions 
shown in Fig.\ref{fig:clusterparameters_fract_wounded}. 
Again the main trends are reproduced: the reconstructed 
cluster size becomes larger 
for peripheral than for central collisions and the width 
parameter $\delta$ increases. 
Discrepancies may be due to the fact, that in reality we have
a mixture of clusters with various sizes and widths
and convolution with wounded nucleons fluctuations gives
in this case different reconstruction results than for
identical clusters.

\vspace{0.3cm}
In the studies of correlations measured over a wide range 
of pseudorapidity, strong long-range effects were found.
There is a long ridge in the correlation with a high-$p_{_{T}}$
trigger particle, clusters found in the 2-particle correlations
are large and unexpectedly wide. The correlation in the azimuthal
angle seems to be determined by the global momentum conservation.
Forward-backward and 2-particle correlations can be at least
qualitatively described by the Wounded Nucleon Model, in which
particles are additionally correlated at large distances
because they are emitted
according to a fragmentation function, which extends over 10 pseudorapidity units. \\[0.3cm]
{\bf Acknowledgments} \\
{\small 
This work was partially supported by U.S. DOE grants 
DE-AC02-98CH10886,
DE-FG02-93ER40802, 
DE-FG02-94ER40818,  
DE-FG02-94ER40865, 
DE-FG02-99ER41099, and
DE-AC02-06CH11357, by U.S. 
NSF grants 9603486, 
0072204,            
and 0245011,        
by Polish MNiSW grant N N202 282234 (2008-2010),
by NSC of Taiwan Contract NSC 89-2112-M-008-024, and
by Hungarian OTKA grant (F 049823).
} \\[0.3cm]

\end{document}